\documentclass[reprint,superscriptaddress,amsmath,amssymb,aps]{revtex4-1}
\usepackage{graphicx}
\usepackage{dcolumn}
\usepackage{bm}
\usepackage{dcolumn}

\begin{document}

\preprint{APS/123-QED}

\title{Physical properties of liquid oxygen under ultrahigh magnetic fields}
\author{T. Nomura}
\email{t.nomura@issp.u-tokyo.ac.jp}
\affiliation{Institute for Solid State Physics, University of Tokyo, Kashiwa, Chiba 277-8581, Japan}
\affiliation{Hochfeld-Magnetlabor Dresden (HLD-EMFL) and W\"urzburg-Dresden Cluster of Excellence ct.qmat, Helmholtz-Zentrum Dresden-Rossendorf, 01328 Dresden, Germany}
\author{A. Ikeda}
\affiliation{Institute for Solid State Physics, University of Tokyo, Kashiwa, Chiba 277-8581, Japan}
\author{M. Gen}
\affiliation{Institute for Solid State Physics, University of Tokyo, Kashiwa, Chiba 277-8581, Japan}
\author{A. Matsuo}
\affiliation{Institute for Solid State Physics, University of Tokyo, Kashiwa, Chiba 277-8581, Japan}
\author{K. Kindo}
\affiliation{Institute for Solid State Physics, University of Tokyo, Kashiwa, Chiba 277-8581, Japan}
\author{Y. Kohama}
\affiliation{Institute for Solid State Physics, University of Tokyo, Kashiwa, Chiba 277-8581, Japan}
\author{Y. H. Matsuda}
\affiliation{Institute for Solid State Physics, University of Tokyo, Kashiwa, Chiba 277-8581, Japan}
\author{S. Zherlitsyn}
\affiliation{Hochfeld-Magnetlabor Dresden (HLD-EMFL) and W\"urzburg-Dresden Cluster of Excellence ct.qmat, Helmholtz-Zentrum Dresden-Rossendorf, 01328 Dresden, Germany}
\author{J. Wosnitza}
\affiliation{Hochfeld-Magnetlabor Dresden (HLD-EMFL) and W\"urzburg-Dresden Cluster of Excellence ct.qmat, Helmholtz-Zentrum Dresden-Rossendorf, 01328 Dresden, Germany}
\affiliation{Institut f\"ur Festk\"orper- und Materialphysik, Technische Universit\"at Dresden, 01062 Dresden, Germany}
\author{H. Tsuda}
\affiliation{National Institute of Advanced Industrial Science and Technology, Tsukuba 305-8586, Japan}
\author{T. C. Kobayashi}
\affiliation{Department of Physics, Okayama University, Okayama 700-8530, Japan}

\date{\today}

\begin{abstract}
We studied the acoustic properties of liquid oxygen up to 90 T by means of ultrasound measurements.
We observed a monotonic decrease of the sound velocity and an asymptotic increase of the sound attenuation when applying magnetic fields.
The unusual attenuation, twenty times as large as the zero-field value, suggests strong fluctuations of the local molecular arrangement.
We point out that the observed fluctuations are related to a liquid-liquid transition or crossover, from a small-magnetization to a large-magnetization liquid, which is characterized by a local-structure rearrangement.
To investigate higher-field properties of liquid oxygen, we performed single-turn-coil experiments up to 180 T by means of the acoustic, dilatometric, magnetic, and optical techniques.
We observed only monotonic changes of these properties, reflecting the absence of the proposed liquid-liquid transition in our experimental conditions.
\end{abstract}
\maketitle

\section{introduction}
Classically, the liquid phase is treated as a homogeneous state of matter where molecules are disordered translationally and orientationally.
In this theoretical framework, the liquid and gas phases have the same symmetry and the density difference is taken as the order parameter to distinguish them.
In the last two decades, discoveries of liquid-liquid transitions (LLTs) have opened a new perspective on the classical understanding of liquids \cite{Poole1997,Tanaka2000,McMillan2007,Gallo2016}.
Two distinguishable states of a liquid indicate that the symmetry of the liquid breaks locally and dynamically.
LLTs have been proposed for many elemental and molecular liquids at high pressure, and are recognized as ubiquitous phenomena \cite{Katayama2000,Cadien2013,Boates2009,Sastry2003,Lorenzen2010,Jara2009,Tanaka2004,Mishima1998,Decremps2018,Ayrinhac2020}.

Up to now, LLTs have been mostly discussed as a function of pressure, $P$, and temperature, $T$ \cite{Anisimov2018}.
For the case of H$_2$O, a low-density liquid (LDL) and a high-density liquid (HDL) are proposed \cite{Gallo2016}.
The frustration between hydrogen-bond (directional) and van der Waals interaction (isotropic) results in two competing states, a tetrahedrally coordinated LDL and close-packed HDL \cite{Tanaka2000,McMillan2007,Gallo2016}.
Pressure tunes the delicate balance between them, and induces a crossover (possibly a phase transition in the supercooled regime).
Here, LDL and HDL are distinguished by density and bond-order parameters; the latter is defined as the fraction of locally favored structures \cite{Tanaka2000}.

When the molecule is magnetic, it is natural to introduce the magnetic field, $H$, as a thermodynamic variable, implying that $H$ can also drive a LLT.
In that case, the magnetization, $M$, is taken as the primary order parameter, and two states of the liquid would be distinguished as small-magnetization liquid (SML) and large-magnetization liquid (LML).
A promising candidate for the magnetic-field-induced LLT is liquid oxygen.
The O$_2$ molecule has a spin $S=1$, and liquid oxygen (90.2--54.4 K) behaves as a paramagnetic liquid with antiferromagnetic (AFM) correlations \cite{Lewis1924,Kratky1975}.
The Curie-Weiss temperature is approximately $\Theta \sim -45$ K, depending on temperature \cite{DeFotis1981,Meier1982,Brodyanskii1989}.
The AFM exchange interaction leads to a dynamical dimerization of O$_2$ molecules with a singlet spin state, that is suggested by magnetic-susceptibility \cite{Kratky1975,Lewis1924}, optical-absorption-spectra \cite{Tsai1969,Uyeda1988,Bhandari1973,Landau1961}, and neutron-scattering data \cite{Fernandez2008,Chahid1993}, as well as by molecular dynamics simulation \cite{Oda2002,Oda2004}.

In fact, an external magnetic field tunes the geometrical alignment of the O$_2$-O$_2$ dimer \cite{Hemert1983,Bussery1993,Bartolomei2008,Obata2013}.
At zero field, the H geometry (rectangular parallel) is the most stable alignment which maximizes the AFM exchange interaction with maximized overlap integral.
When the magnetic moment is fully polarized by external fields, the O$_2$-O$_2$ dimer prefers X (crossed) or S (canted) geometries to minimize the overlap integral.
Such a field-induced molecular rearrangement has been observed in the solid oxygen $\alpha$-$\theta$ phase transition at around 100 T \cite{Nomura2014,Nomura2015,Nomura2017PD}.
The same mechanism would be relevant also for a LLT of oxygen where the favored molecular alignment (H, X, S, distinguished by the bond-order parameter) changes by magnetic field.

In this paper, we present the ultrasonic properties of liquid oxygen up to 90 T by using nondestructive pulsed magnets \cite{Nomura_arxiv}.
The obtained results suggest that magnetic fields induce strong fluctuations of the molecular arrangement.
To extend the measurements up to higher magnetic fields, we employed single-turn-coils (STCs), that is a semi-destructive pulse-field-generation technique.
We investigated the acoustic, dilatometric, magnetic, and optical properties of liquid oxygen up to 180 T.
We discuss the obtained results in the context of the proposed magnetic-field-induced LLT.

\section{Experiment}
\subsection{Ultrasound measurement with nondestructive magnets}
\begin{figure}[ptb]
\centering
\includegraphics[width=8.6cm]{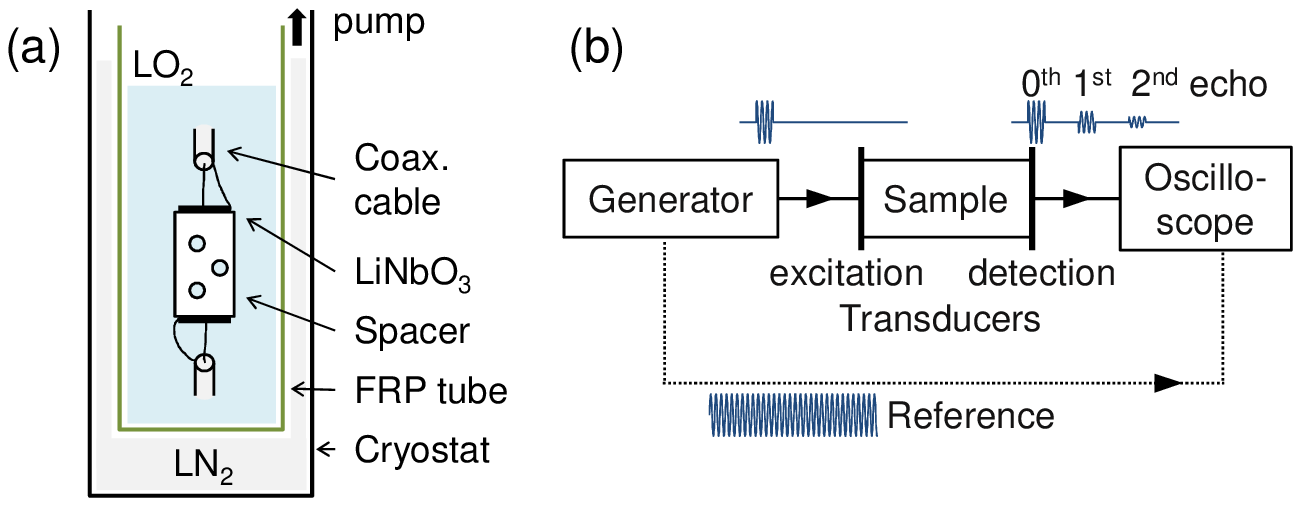}
\caption{\label{fig:exp}
(a) Experimental configuration of the ultrasound measurement for liquid oxygen with nondestructive pulsed magnets.
(b) Block diagram of the ultrasound pulse-echo technique.
}
\end{figure}

We used an ultrasound pulse-echo technique with digital homodyne detection to study the sound velocity, $v$, and acoustic-attenuation coefficient, $\alpha$.
We show a scheme of the experiment in Figs. \ref{fig:exp}(a) and \ref{fig:exp}(b).
We placed two LiNbO$_3$ transducers (Y-36$^\mathrm{o}$ cut, fundamental resonance at 20--40 MHz) parallel to each other by using a plastic spacer (with typical sample length $L=5$ mm).
The sample space was filled by pure oxygen gas (99.999 \%).
On cooling, liquid oxygen at vapor pressure was condensed at the bottom of a fiber-reinforced plastic (FRP) tube.
Condensed oxygen was automatically filled in the plastic spacer through holes.
Radio frequency (RF, 20--210 MHz) acoustic-wave pulses were excited and detected by the transducers.
The echo signal and reference RF were recorded by a digital oscilloscope.
The phase and amplitude of the 0th echo were detected by a digital homodyne technique and converted to the relative change of $v$ and $\alpha$ as,
\begin{equation}
\Delta v/v=\Delta \Phi/\Phi=\Delta \Phi/(2\pi f \tau_0),
\label{eq:vel_phase}
\end{equation}
\begin{equation}
\Delta \alpha=-\mathrm{ln}(A/A_0)/L.
\label{eq:att_amp}
\end{equation}
Here, $\Phi$ is the phase of the detected echo, $f$ is the ultrasound frequency, $\tau_0$ is the travel time of the sound for $L$, and $A/A_0$ is the amplitude normalized to the zero-field value.
The sound velocity at zero field is calibrated by the value from literature ($v \sim 1000$ m/s at 77 K) \cite{Itterbeek1962,Dael1966,Clouter1973}.
$v$ does not depend on $f$ below 3.6 GHz \cite{Clouter1973}.

Magnetic fields below 60 T were generated by a pulsed magnet with pulse duration of 150 ms.
The 90 T pulse was generated by a dual-coil-pulse magnet (with the field profile shown in the inset of Fig. \ref{fig:attenuation}) \cite{Zherlitsyn2013}.
The magnetic field was measured by a pickup coil placed near the plastic spacer.
The magnetic field was aligned parallel to the ultrasound propagation direction.
The temperature immediately before the pulse was measured by a RuO$_2$ resistance thermometer placed near the sample cell.

\subsection{Single-turn-coil experiments} 
To extend the measurement to higher magnetic fields, STC systems at ISSP, Kashiwa were used \cite{Miura03}.
A vertical-type STC was used for ultrasound, magnetovolume-effect, and magnetization measurements.
Optical absorption spectroscopy was performed at a horizontal-type STC system.
Magnetic fields up to 180 T were generated with a duration of $\sim$7~$\mu$s.
A pickup coil located near the sample detected the inductive voltage $dB/dt(t)$, and the magnetic field $B(t)$ was obtained by numerical integration.

\begin{figure}[ptb]
\centering
\includegraphics[width=8.2cm]{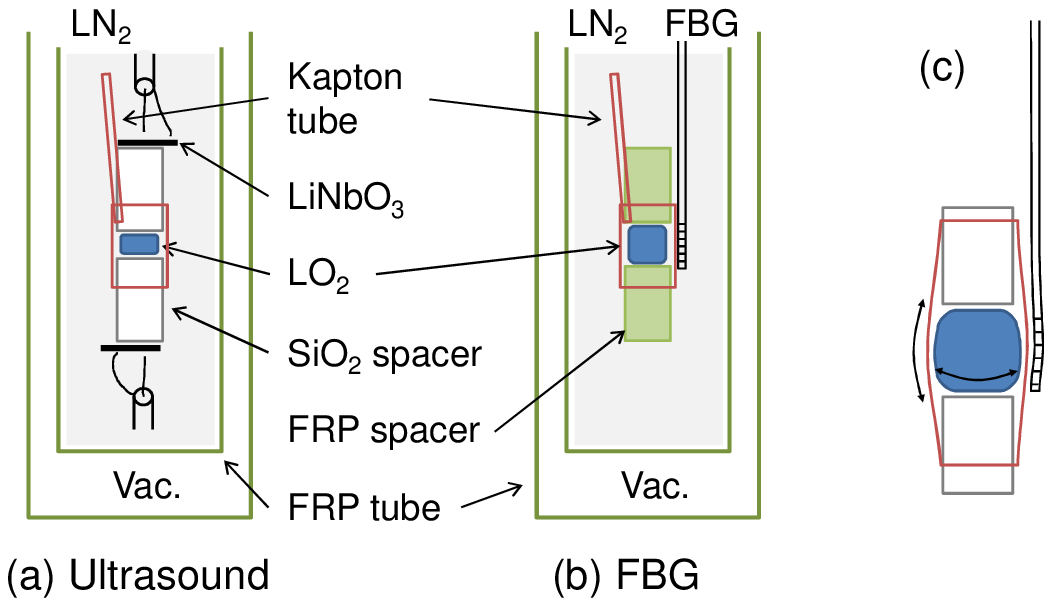}
\caption{\label{fig:Kapton}
Experimental configurations for the (a) ultrasound and (b) magnetovolume-effect measurements with the vertical STC system.
A fiber-Bragg grating (FBG) is attached on the surface of the Kapton tube filled with liquid oxygen.
(c) Axial and radial deformations of the Kapton cell due to the magnetovolume effect of liquid oxygen.
}
\end{figure}

Ultrasound measurement was performed using a continuous wave (CW) technique \cite{Nomura2021RSI}.
One transducer (LiNbO$_3$ Y-36$^\mathrm{o}$ cut) was continuously excited with the resonant frequency around 40 MHz, and another transducer detected the transmitted ultrasonic waves.
The detected signal was analyzed by the numerical homodyne technique and the relative change of the sound velocity was obtained using Eq. (\ref{eq:vel_phase}).
The sample cell was composed of two SiO$_2$ glass cylinders and a Kapton tube of 3~mm diameter [Fig. \ref{fig:Kapton}(a)].
All the parts were connected by low temperature glue (Nitofix, SK-229).
The sample length (the gap between the two glass cylinders) was $\sim0.3$ mm.
Oxygen gas was introduced to the gap between the glass cylinders through a thin Kapton tube, and condensed at 77 K in a liquid-N$_2$ bath cryostat.
Two transducers were attached on the outer surfaces of the glass cylinders.
The two cylinders worked as acoustic delay lines for ultrasound measurement.
For details, see Ref. \cite{Nomura2021RSI}.

The magnetovolume effect was measured by a fiber-Bragg grating (FBG) attached on the side surface of the sample cell.
The schematic setup, which was derived from the ultrasound measurement, is shown in Fig. \ref{fig:Kapton}(b).
The sample cell was composed of two FRP cylinders and a Kapton tube of 2~mm diameter.
The gap between the two cylinders (sample length) was 2.0 mm.
We call this cell as the $\Phi 2$ cell.
We also employed the larger $\Phi 3$ cell (diameter = length = 3 mm) for comparison.
Oxygen gas was introduced through a Kapton tube (0.6 mm diameter, $\sim50$ mm long).
The thin and long Kapton tube allowed gaseous oxygen to pass, but did not allow liquid oxygen to escape from the sample cell on a microsecond timescale.
A preceding study showed that the volume of liquid oxygen increases by 0.022~\% at 8 T \cite{Uyeda1987}.
When liquid oxygen expands by applying pulsed magnetic fields, the Kapton cell is expected to deform along the axial and radial directions.
The attached FBG detected the strain at the surface of the sample cell.
The strain of the FBG was observed as a shift of the optical reflection spectrum, which was monitored as an intensity change by using an optical filter.
To check the quantitative reliability for each setup, we first performed test measurements by using non-destructive magnets and used the same setup in STC experiments.
For details of the FBG technique with STCs, see Ref. \cite{Ikeda2017RSI}.

The magnetization measurement was performed by using a parallel-type twin pickup coil \cite{12JPSJ_Takeyama,Nomura2015}.
We have mounted two oppositely wound pickup coils (16 x 2 turns) on a Kapton tube with 1 mm diameter.
Liquid oxygen was condensed in a Kapton tube of 0.7 mm and inserted into one side of the compensated pickup coil.
The pickup coil detected the inductive voltage due to the external field and the magnetic moment of the sample, where the former was compensated while the latter was not.
The sample setup was cooled down to 55 K by using a $^4$He-gas-flow cryostat.
We note that a liquid-N$_2$ bath cryostat could not be used for this measurement.
When the twin pickup coil and the sample tube mechanically touched, even via liquid nitrogen, the strain of liquid oxygen was transfered to the pickup coil.
In such cases, the deformed pickup coil could not compensate the induced voltage.
Therefore, a slight gap between the sample tube and pickup coil (0.7 and 1.0 mm diameter) was quite important.
For details and schematic figures, see the Supplemental Material of Ref. \cite{Nomura2015}.

The magnetic-field dependence of the optical transmission was measured using a high-speed streak camera.
A $^4$He-flow-type cryostat made of Bakelite was used to cool the sample. 
The sample space inside the cryostat was sandwiched by two optical fibers and sealed by Nitofix.
Oxygen gas was introduced through a Kapton tube and condensed.
The temperature was monitored by a type-E thermocouple build in the sample space.
Incident light was generated by a Xe arc flash lamp and guided to the sample via an optical fiber.
The transmitted light was collected by the other optical fiber and transfered to the streak camera with a polychromator.
For details and schematic figures, see Refs. \cite{Nomura2013,Nomura2014}.

\section{Results and discussion}
\subsection{Acoustic properties up to 90 T}

\begin{figure}[ptb]
\centering
\includegraphics[width=7.4cm]{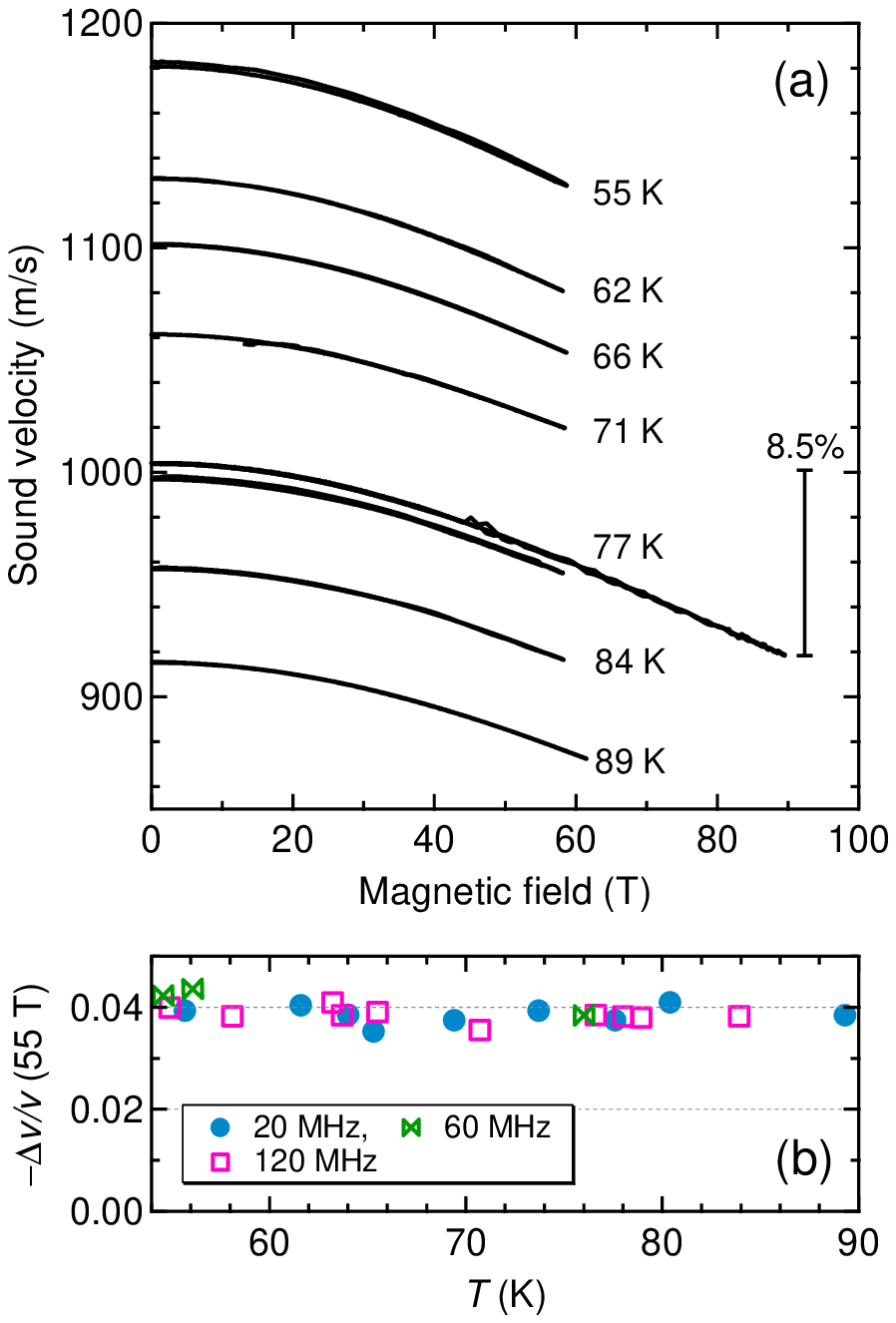}
\caption{\label{fig:velocity}
(a) Sound velocity of liquid oxygen as a function of magnetic field.
Temperatures are denoted at each curve.
(b) Relative change of the sound velocity at 55 T as a function of temperature for each ultrasound frequency.
}
\end{figure}

First, we present the results of the ultrasound measurement by using non-destructive magnets.
Figure~\ref{fig:velocity}(a) shows the sound velocity as a function of magnetic field at various temperatures.
In the whole temperature range, a monotonic decrease of $v$ is observed without any hysteresis.
The reversible results indicate that any	 temperature change (e.g., magnetocaloric effect) is negligible in this experimental setup (see also Ref. \cite{Nomura2017MCE}).
At 90~T, $v$ decreases by 8.5 \% without any sign of saturation.
Figure \ref{fig:velocity}(b) shows the temperature and frequency dependence of the relative change of the sound velocity $\Delta v/v$ at 55 T.
Within our measurement accuracy, $\Delta v/v$ does not depend on $T$ and $f$.

The decrease of the sound velocity can partially be attributed to the magnetovolume effect of liquid oxygen \cite{Uyeda1987}.
At zero field, the most stable molecular configuration is H-type geometry, that results in antiferromagnetic (AFM) correlations.
When the magnetic moments of the O$_2$ molecules are polarized by external fields, the O$_2$-O$_2$ distance tends to be larger to suppress the AFM coupling.
This exchange-striction mechanism gives rise to the volume expansion of liquid oxygen under magnetic fields.
The decrease of the density, $\rho$, indicates a decrease of intermolecular binding energy, which originates from van der Waals and exchange interactions.
As a result, the bulk modulus of liquid oxygen $K$ decreases, leading to a decrease of sound velocity as $v=\sqrt{K/\rho}$.
The change of the bulk modulus $|\Delta K/K|$ is considered to be larger than that of the density $|\Delta \rho / \rho|$.

\begin{figure}[ptb]
\centering
\includegraphics[width=8.0cm]{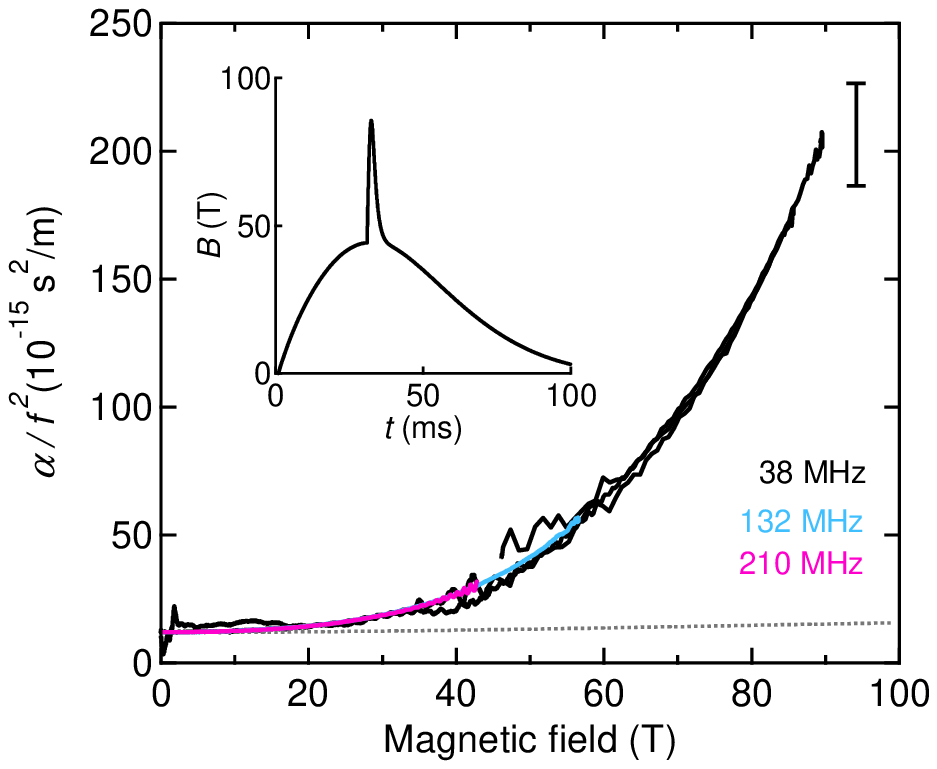}
\caption{\label{fig:attenuation}
Normalized acoustic-attenuation coefficient $\alpha/f^2$ of liquid oxygen at 77 K as a function of magnetic field.
The error bar at 90 T is shown.
The dotted line shows the classically expected behavior based on Eq. (\ref{eq:expected_att}).
The inset shows the time dependence of the pulsed magnetic field.
}
\end{figure}

Figure \ref{fig:attenuation} shows the acoustic attenuation coefficient at 77 K up to 90 T.
Here, we plot $\alpha$ normalized by the square of the ultrasound frequency, $\alpha/f^2$ \cite{Bhatia}, which leads to one universal curve for the different ultrasound frequencies.
In this paper, we use the unit $10^{-15}$s$^2$/m for the normalized attenuation.
Up to 90 T, $\alpha/f^2$ shows  a monotonous increase and reaches a value above 200, which is twenty times as large as the zero-field value.
Empirically, $\alpha/f^2$ is known to be approximately 10 for the simple non-associated liquids including oxygen at zero field (Table \ref{tab:table1}) \cite{Victor1970,Galt1948,Hall1948}.
This evidences that the observed value of 200 is extraordinarily large.
$\alpha/f^2$ seems still to increase strongly towards higher magnetic fields.

\begin{table}
\caption{\label{tab:table1}Normalized acoustic-attenuation coefficients $\alpha/f^2$ for simple molecular liquids.}
\begin{ruledtabular}
\begin{tabular}{lD{.}{.}{5}D{.}{.}{5}r}
Liquid & T\ (\mathrm{K}) & \alpha/f^2\ (10^{-15}\ \mathrm{s}^2/\mathrm{m}) & Ref. \\
\hline
O$_2$ & 77 & 11 & \cite{Victor1970} \footnote{zero-field value}\\
N$_2$ & 73.9 & 10.6 & \cite{Galt1948} \\
H$_2$ & 17 & 5.6 & \cite{Galt1948} \\
Ar & 85.2 & 10.1 & \cite{Galt1948} \\
H$_2$O & 353 & 10.6 & \cite{Hall1948}\\
H$_2$O & 278 & 61.4 & \cite{Hall1948}\\
\end{tabular}
\end{ruledtabular}
\end{table}

Next, we discuss the origin of the observed sound attenuation.
The classical sound attenuation due to viscosity and heat conduction \cite{Victor1970,Bhatia} is described by 
\begin{equation}
\frac{\alpha}{f^2}=\frac{2\pi^2}{\rho v^3}\left\{ \frac{4}{3}\eta_s+\eta_v+\frac{(\gamma-1)\kappa}{c_p} \right\}.
\label{eq:norm_att}
\end{equation}
Here, $\eta_s$ is the shear viscosity, $\eta_v$ is the bulk viscosity, $\kappa$ is the thermal conductivity, $c_p$ is the specific heat at constant pressure.
$\gamma = c_p/c_v$ is the ratio of specific heats.
Typical values are summarized in Table \ref{tab:table2} \cite{Victor1970}.
The acoustic attenuation dominantly originates from the viscosity ($\eta_s$ and $\eta_v$), and the contribution from the heat conduction is around 10~\% \cite{Victor1970}.

\begin{table*}
\caption{\label{tab:table2} Typical parameters of liquid oxygen at zero field \cite{Victor1970} and their contributions to the acoustic attenuation.}
\begin{ruledtabular}
\begin{tabular}{ccccccc}
$T$ & $\eta_s$ & $\eta_v$ & $\gamma$ & $\kappa$ & $c_p$  & 4/3$\eta_s$ : $\eta_v$ : ($\gamma-1$)$\kappa$/$c_p$ \\
(K) & (10$^{-3}$ Pa$\cdot$s) & (10$^{-3}$ Pa$\cdot$s) &  & (J/m$\cdot$s$\cdot$K) & (J/kg$\cdot$K) & \\
\hline
80.1 $\pm$ 0.2 & 0.245 & 0.212 & 1.687 & 0.163 & 1680 & 0.54 : 0.35 : 0.11\\
75.4 $\pm$ 0.2 & 0.282 & 0.279 & 1.678 & 0.167 & 1670 & 0.52 : 0.39 : 0.09\\
\end{tabular}
\end{ruledtabular}
\end{table*}

Assuming that the value inside the curly bracket in Eq. (\ref{eq:norm_att}) is field independent, the relative change of $\alpha$ is given by
\begin{equation}
\Delta\alpha/\alpha=-\Delta\rho/\rho-3\Delta v/v.
\label{eq:expected_att}
\end{equation}
The density decreases by 0.02~\% at 8~T \cite{Uyeda1987}, and a quadratic extrapolation gives $-2.8$~\% at 90~T.
Using the measured $\Delta v/v$ [Fig.~\ref{fig:velocity} (a)], the field dependence of the sound attenuation can be calculated as the dotted line in Fig.~\ref{fig:attenuation}.
$\Delta \alpha/f^2$ is at most 3, which is negligibly small compared to the observed value.
Let us comment on our assumption of the field independence of the curly bracket of Eq. (\ref{eq:norm_att}).
The dominant contribution of this term comes from $\eta_s$ and $\eta_v$.
The viscosity measurement of liquid oxygen suggests that $\eta_s$ decreases by 5 \% at 50~T \cite{Comment_visc}, that even reduces the estimated change of $\alpha$.
When the shear viscosity decreases, the bulk viscosity should also decrease because both are related to intermolecular interactions.
The contribution from the last term due to heat conduction ($\gamma-1$)$\kappa$/$c_p$ is relatively small, and would not change the discussion above.
Thus, the classical attenuation mechanisms due to viscosity and heat conduction cannot account for the observed strong acoustic attenuation.

As possible reasons for the excessive acoustic attenuation, we consider (i) relaxation effect and (ii) a phase transition (including critical phenomena) \cite{Bhatia}.
The effect of relaxation is observed when the ultrasound frequency is close to the relaxation dynamics of the molecules.
As an estimate for that, the so-called Lucas relaxation frequency is used \cite{Fleury1969},
\begin{equation}
f_\mathrm{L}=\frac{\rho v^2}{4/3\eta_s+\eta_v}\sim 2\ \mathrm{THz}.
\label{eq:Lucas}
\end{equation}
The parameters are taken from Table \ref{tab:table2} \cite{Victor1970}.
$f_\mathrm{L}$ is consistent with the relaxation time obtained from inelastic neutron scattering \cite{Fernandez2008} and is four orders of magnitude larger than the ultrasound frequencies used in this study.
Moreover, $\Delta v/v$ and $\Delta \alpha/f^2$ do not depend on the ultrasound frequencies (Figs. 2 and 3), indicating that the dispersion is negligible in this frequency range.
Therefore, we can exclude the effect of relaxation.

As a possible explanation for the excessive acoustic attenuation we, therefore, propose a precursor of a phase transition (including critical phenomena) from a SML to a LML.
The locally favored structure changes from H- to X- or S-type geometry by applying a magnetic field \cite{Hemert1983,Bussery1993,Bartolomei2008,Obata2013}, that results in a phase transition for solid oxygen at $\sim$100 T \cite{Nomura2014,Nomura2015,Nomura2017PD}.
Near the phase boundary (or critical line), the local molecular arrangement is more degenerated.
The fluctuation of the local structure and density results in the strong acoustic attenuation and the decrease of sound velocity.
This is generally observed near a gas-liquid critical point \cite{Garland1970,Thoen1974,Garland1970_2}.
For the case of liquid H$_2$O, the increase of $\Delta \alpha/f^2$ by lowering the temperature (see Table \ref{tab:table1}) is explained in the context of two competing local molecular arrangements \cite{Hall1948}.
The observed increase of $\Delta \alpha/f^2$ could also be attributed to field-induced local-structure fluctuations, namely, a precursor of the field-induced LLT.

When the fluctuations are suppressed at extremely high fields ($\sim$300 T), $\alpha/f^2$ should show reasonable values in the framework of the classical theory ($\sim$10).
Therefore, a maximum of $\alpha/f^2$ is expected at higher magnetic fields, where the LLT or crossover takes place.
Thermodynamic analysis suggests that a second-order phase boundary is allowed for the vector order parameter $\Delta \bf M$, but not for the scalar order parameter $\Delta \rho$ \cite{Anisimov2018}.
In this sense, the magnetic-field-induced LLT might show qualitatively different behavior from the pressure-induced LLT.
Experiments at higher fields are needed to prove the existence of the magnetic-field-induced LLT.

In the following sections, we present the results obtained in the higher field region up to 180 T.
Even in this field range, we could not obtain results indicating a magnetic-field-induced LLT.
Because of the very short field duration and strong electromagnetic noise of the STC, the obtained results may be partially less reliable.
Therefore, in the following sections, we particularly pay attention on the reliability and reproducibility of the results.

\subsection{Ultrasound velocity in STC experiments}

\begin{figure}[ptb]
\centering
\includegraphics[width=8.0cm]{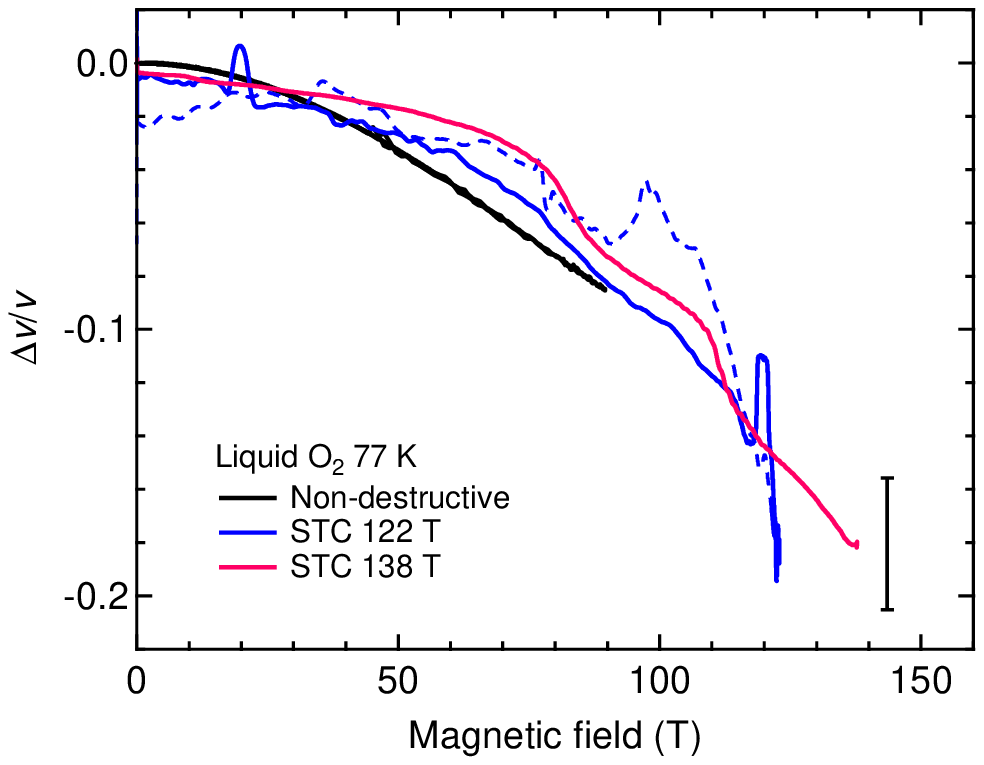}
\caption{\label{fig:us_stc}
Relative changes of the sound velocity of liquid oxygen at 77 K as a function of magnetic field.
The result for the field-down sweep is shown by the dashed line.
The estimated error bar at maximum field is shown.
}
\end{figure}

Here, we present the ultrasound results by using the continuous technique and the STC system \cite{Nomura2021RSI}.
With this technique, the obtained acoustic attenuation is less reliable because of the acoustic interference and strong noise from the STCs.
Thus, we only present the results for the sound velocity.
Figure \ref{fig:us_stc} shows representative results up to 138 and 122 T together with the result up to 90 T obtained by using the non-destructive magnet.
The error bar is estimated from five repeated experiments up to 100--130 T, that are qualitatively consistent.
For the pulse up to 138 T, we show only data for the up sweep, since the signal was greatly disturbed after reaching the maximum field probably due to broken transducers.
These results show a continuous decrease of $\Delta v/v$ by $\sim$20~\% at 138 T, from 1000 to 800 m/s, indicating that no LLT occurs up to this field.
Indeed, this sound velocity is already smaller than the value at the boiling point (90~K) at zero field.
The decrease of sound velocity does not show any tendency of saturation.

We note that the sound velocity of a material in the liquid state cannot be smaller than that in the gaseous state.
The sound velocity of gaseous O$_2$ is $\sim$300 m/s \cite{GasO2}.
Therefore, the decrease of the sound velocity cannot continue to arbitrarily high fields.
Instead, it should show a minimum related to the LLT or a saturation when the magnetization saturates.
Because of the challenging experimental setup, the ultrasound measurements were successful only up to the field range shown.
In the following sections, the experimental results by using other probes are presented.

\subsection{Magnetovolume effect in STC experiments}
The volume (density) change is an important measure to detect pressure-induced LLTs \cite{Poole1997,Tanaka2000,McMillan2007,Gallo2016}.
For the case of liquid oxygen, the magnetic correlation is strongly coupled to the molecular geometry.
Therefore, the magnetovolume effect would also detect magnetic-field-induced LLTs.
In this paper, we present the results up to 50 T (non-destructive magnets) and up to 183 T (STC) at 77~K.

\begin{figure}[ptb]
\centering
\includegraphics[width=8.6cm]{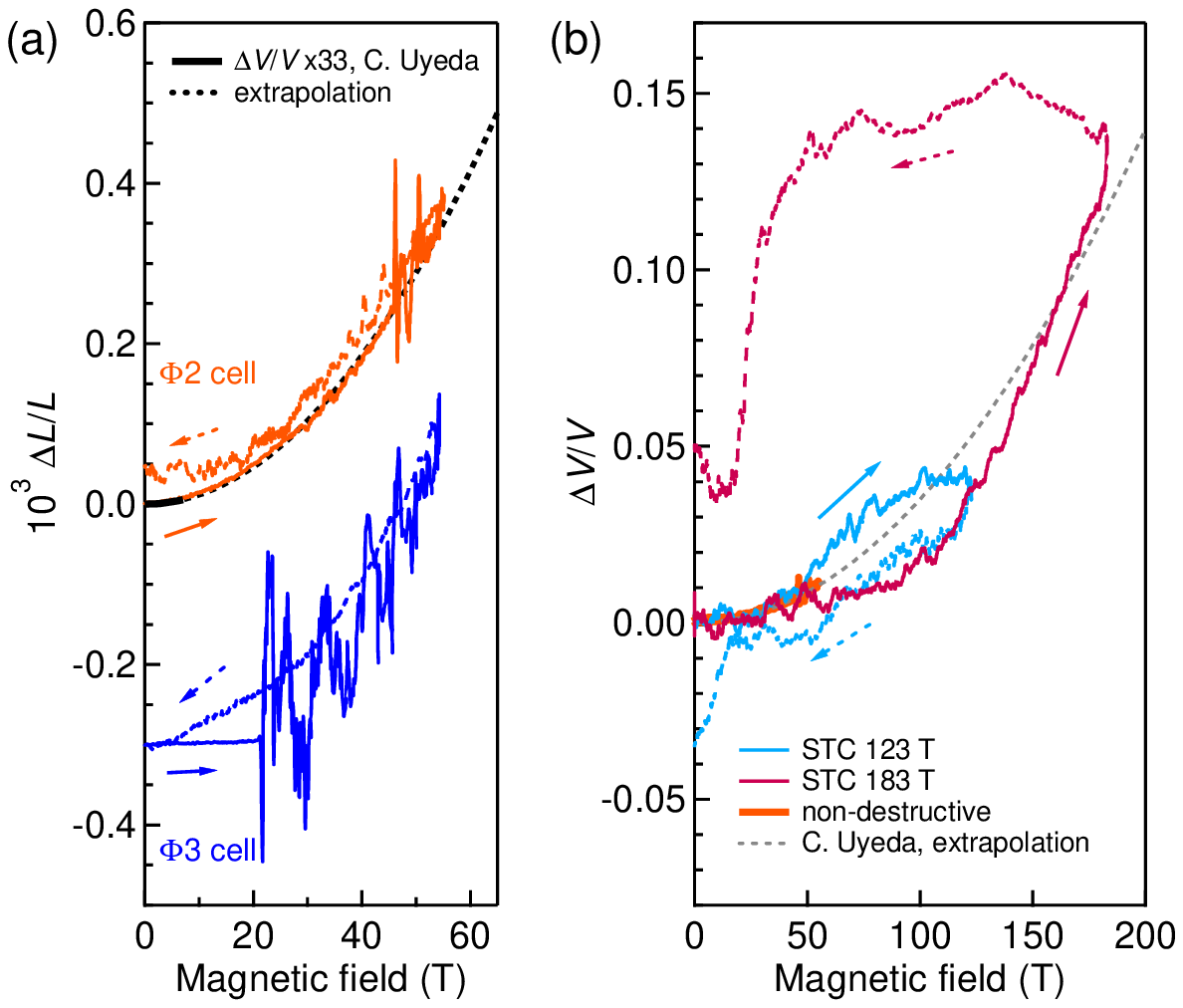}
\caption{\label{fig:FBG}
Results of magnetovolume-effect measurements at 77 K.
(a) $\Delta L/L$ determined from the deformation of FBGs bonded on the surface of  sample cells. 
The result reported by C. Uyeda \cite{Uyeda1987} is multiplied by 33, and its extrapolation is shown for comparison.  
(b) Relative volume change of liquid oxygen $\Delta V/V$ estimated from $\Delta L/L$. 
Results for the field-up (-down) sweeps are shown by solid (dashed) lines.
}
\end{figure}

First, the results up to 50 T are presented in Fig. \ref{fig:FBG}(a).
Here, the deformation of the FBG $\Delta L/L$, which is related to the volume change of liquid oxygen, is shown.
The obtained $\Delta L/L$ curves are nearly reversible and increase quadratically as a function of external field.
The slight difference for the field-up sweep (solid line) and field-down sweep (dashed line) is probably due to some movement of the liquid in the cell.
In the timescale of the non-destructive magnet (36 ms duration), liquid oxygen may partially move out from the cell through the capillary [Fig. \ref{fig:Kapton}(b)].
We also noticed that the hysteresis depends on the size of the sample cell.
In Fig. \ref{fig:FBG}(a), the blue curve is the result using the $\Phi3$ cell, while all other curves in Fig.~\ref{fig:FBG} are obtained by using the $\Phi2$ cell.
For the $\Phi3$ cell, $\Delta L/L$ changes only a few ms after the field pulse starts.
This indicates that the stress inside the cell is not effectively transfered to the FBG, probably because of inhomogeneous compression in the cell.
By using the $\Phi2$ cell, the response delay becomes negligible and the noise level also becomes smaller.

Next, a conversion relation between the deformation $\Delta L/L$ measured using a FBG and the volume expansion of liquid oxygen $\Delta V/V$ is considered.
Note, that this relation depends on the manner of deformation (axial, radial, and their combination) and the coupling strength between the FBG and Kapton cell.
Therefore, this conversion relation needs to be calibrated for each experimental setting.
In Fig. \ref{fig:FBG}(a), the experimental $\Delta V/V$ up to 8 T \cite{Uyeda1987}, which is proportional to $B^2$, is shown.
This reported curve is normalized and quadratically extrapolated to compare with $\Delta L/L$.
Nonlinear terms [$(\Delta V/V)^{1/2}$ or $(\Delta V/V)^{2}$] are not necessary to reproduce the results.
Therefore, at least in this field range, $\Delta V/V$ is proportional to $\Delta L/L$.
By assuming that this relation holds as well at higher fields, the experimental results are converted to $\Delta V/V$.

Figure \ref{fig:FBG}(b) shows the estimated $\Delta V/V$ of liquid oxygen up to 183 T.
Although there is no guarantee of $\Delta V/V \propto \Delta L/L$ in this field range, the qualitative change of $\Delta V/V$ is considered to be reliable. 
Both results, up to 123 T and 183 T, roughly follow the extrapolated curve of $\Delta V/V$ \cite{Uyeda1987}.
This indicates that the magnetization of liquid oxygen is not saturated even in this field range.
At the peak field of 183 T, the result became unstable probably because the glue between the FBG and sample cell was damaged.
These results suggest that there is no LLT even up to 183 T.

This is a surprising outcome because the Curie-Weiss temperature of liquid oxygen is only $\Theta \sim -45$ K \cite{Meier1982,DeFotis1981,Brodyanskii1989}.
The magnetic field of 183 T should be more than sufficient to break the AFM coupling between the O$_2$ molecules.
To explain the unexpectedly large saturation field (above 180 T), our experimental condition, namely isobaric or isometric conditions, might be important.
Our sample cells for the STC experiments are designed to keep the liquid oxygen in a closed space.
However, even for the FBG cell, the large volume change of the order of a few percents is not acceptable.
Indeed, after the STC experiment, the sample cells were often broken because of the pressure increase.
Taking into account the estimated volume increase of 10~\%, our experiment is considered to be close to the isometric condition.
In this case, the increasing pressure can enhance the magnetic interactions compared to the isobaric condition.

\begin{figure}[ptb]
\centering
\includegraphics[width=8.2cm]{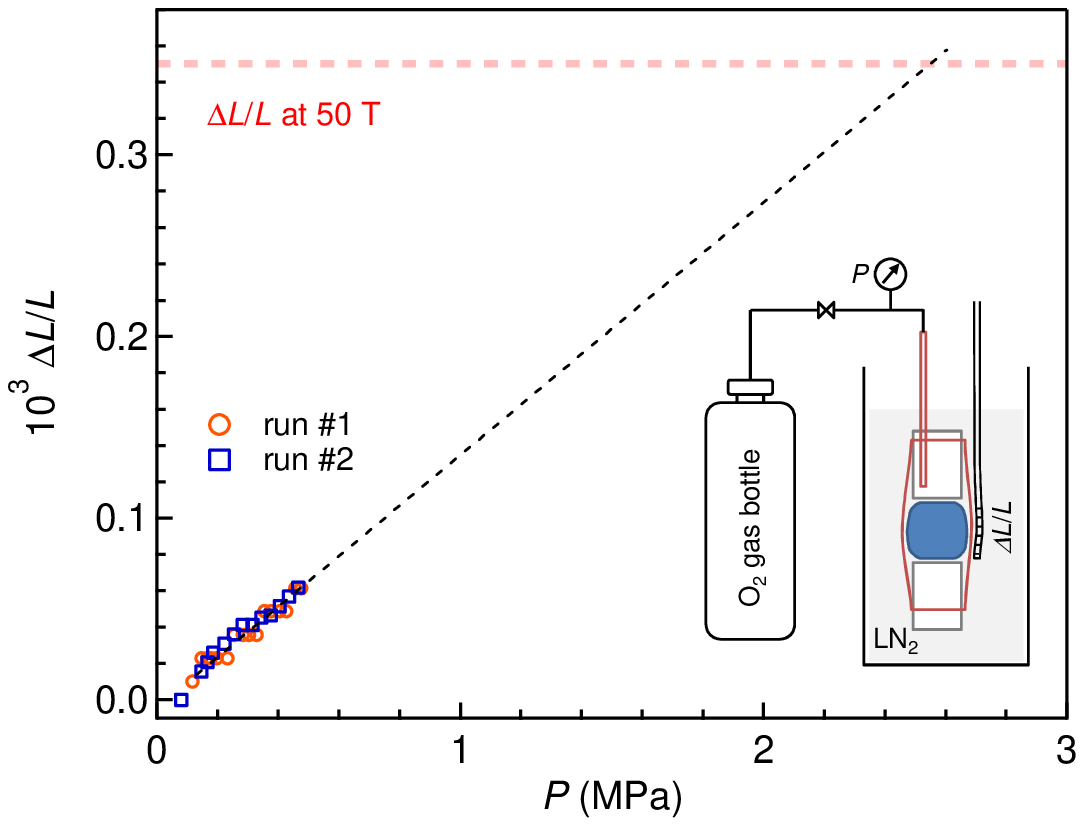}
\caption{\label{fig:P_calib}
$\Delta L/L$, measured by use of a FBG, as a function of pressure $P$ inside the cell at 77 K.
The schematic experimental setting is shown in the inset.
$\Delta L/L$ at 50 T [see Fig. \ref{fig:FBG} (a)] and extrapolation of $\Delta L/L(P)$ are shown by dashed lines.
}
\end{figure}

Next, we qualitatively estimate the pressure, $P$, inside the cell.
To obtain the relation between $P$ and $\Delta L/L$ for the FBG cell, we performed a zero-field calibration.
Here, liquid oxygen is condensed in the FBG cell which is kept in a liquid nitrogen bath.
By gradually introducing O$_2$ gas, $\Delta L/L$ as a function of $P$ is obtained (Fig.~\ref{fig:P_calib}).
Experiments above 0.5 MPa were not successful because of the limited durability of the glue.
The obtained $\Delta L/L(P)$ is linear in the measured pressure range.
By extrapolating $\Delta L/L$ to $0.33 \times 10^3$, the pressure at 50 T is estimated to be 2.5 MPa (intersection of the two dashed lines in Fig.~\ref{fig:P_calib}).
This pressure would destroy the cell under static conditions.
Therefore, it is understandable that the results for fields up to 183 T become noisy although the pressure increase occurs in a few $\mu$s.
As long as we use closed cells as liquid-oxygen containers, such pressure increase cannot be avoided.
This situation applies to ultrasound, magnetization, as well as to optical measurements as discussed later.

\subsection{Magnetization in STC experiments}
\begin{figure}[ptb]
\centering
\includegraphics[width=8.2cm]{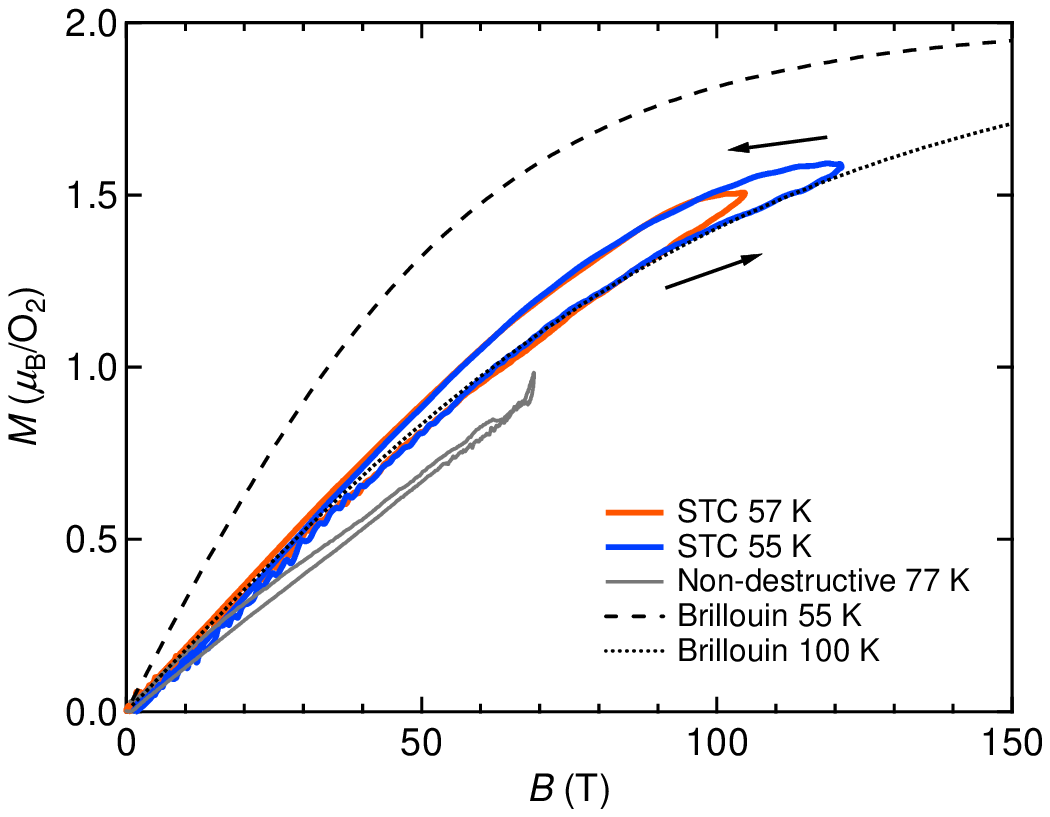}
\caption{\label{fig:mag}
Magnetization curves of liquid oxygen obtained by using STC and non-destructive magnets.
Theoretical curves with and without magnetic interaction are shown by dotted and dashed curves, respectively.
}
\end{figure}

For observing the LLT from a SML to a LML, magnetization measurement are the most straightforward technique because the magnetization difference is the primary order parameter.
Here, we present the results of such magnetization measurements at low temperatures.
Figure \ref{fig:mag} shows the magnetization of liquid oxygen obtained by using STC and non-destructive magnets.
As discussed in Sec. II-B, the STC results were obtained in a He-gas-flow cryostat.
The result up to 70 T, obtained by using a nondestructive magnet, show negligible hysteresis, while the STC results show slight hysteresis.
The latter is considered to be extrinsic, caused by the deformation of the sample cell by the magnetovolume effect of liquid oxygen.
In other words, the amount of sample coupled to the pickup coil (filling factor) might have changed during the magnetic-field application.

For quantitative discussions, the Brillouin function 
\begin{equation}
\frac{M}{g\mu_\mathrm{B}}=\frac{2S+1}{2S}\mathrm{coth}(\frac{2S+1}{2S}x)-\frac{1}{2S}\mathrm{coth}(\frac{1}{2S}x),
\label{eq:bri1}
\end{equation}
\begin{equation}
x=g\mu_\mathrm{B}SH/k_\mathrm{B}T,
\label{eq:x1}
\end{equation}
with $S=1$ and $T=55$ K is shown by the dashed curve (Fig. \ref{fig:mag}).
Here, $g=2$ is the $g$-factor, $\mu_\mathrm{B}$ is the Bohr magneton, and $k_\mathrm{B}$ is the Boltzmann constant.
This curve corresponds to the average magnetization of the O$_2$ molecules without any interaction.
The difference between the experimental and the Brillouin curves reflects the strong AFM correlations in liquid oxygen.
This AFM correlation effect can be treated by a modified Brillouin function introducing an effective temperature $T^*=T-\Theta$, where $\Theta$ is the Curie-Weiss temperature.
By using $\Theta=-45$ K \cite{Meier1982,DeFotis1981,Brodyanskii1989}, we obtain the dotted magnetization curve in Fig.~\ref{fig:mag} ($T^*=100$ K).
In the LLT scenario, the dashed (dotted) curve would correspond to the magnetization of LML (SML), if the AFM interaction is completely suppressed in the LML phase.
The experimental magnetization follows the dotted curve up to 120 T, suggesting no LLT in this field range.
At 120 T, the magnetization is only 75 \% of the saturation value, which is consistent with the magnetovolume-effect results showing no tendency of saturation.

\subsection{Magneto-optical properties in STC experiments}
\begin{figure*}[ptb]
\centering
\includegraphics[width=18cm]{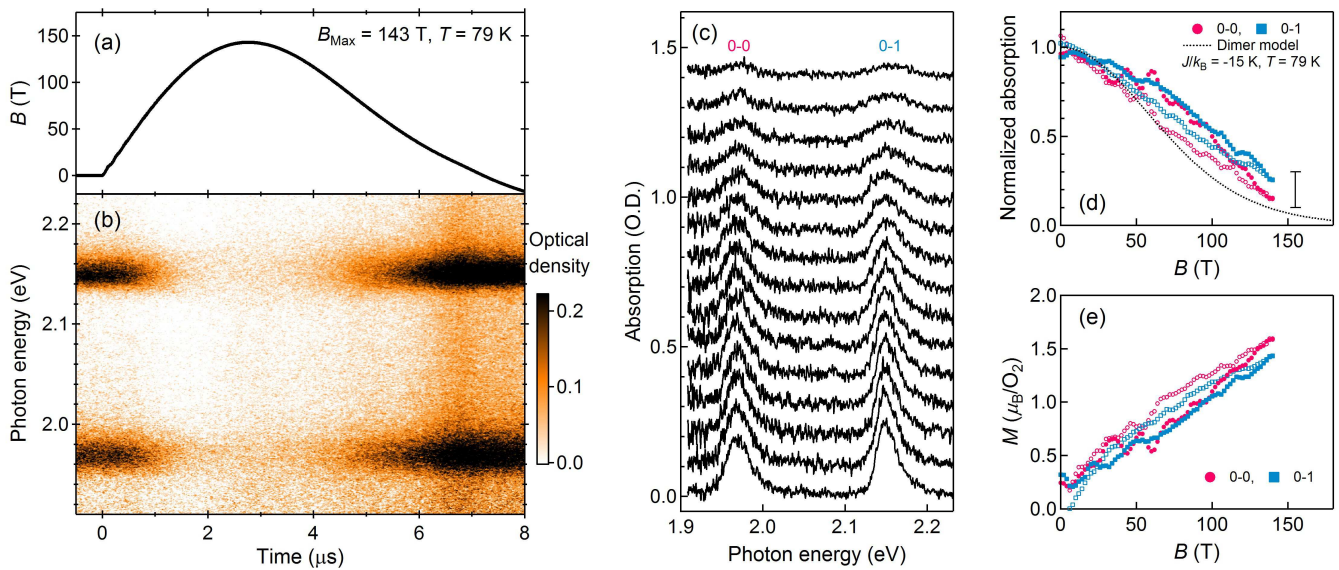}
\caption{\label{fig:opt}
Typical results of magneto-optical measurements on liquid oxygen.
The maximum field and temperature are 143 T and 79 K, respectively.
(a) Magnetic field and (b) absorption spectra as a function of time.
(c) Optical absorption spectra of liquid oxygen for magnetic fields from 0 T (bottom) to 140 T (top) in steps of 10 T.
We show the up-sweep spectra shifted vertically for clarity.
(d) Normalized absorption intensity of the (0-0) and (0-1) peaks as a function of magnetic field.
Results for the up and down sweeps are shown by filled and open symbols, respectively.
The calculated absorption intensity, based on Eq. (\ref{eq:absAF}), is shown by the dotted curve.
(e) Converted magnetization curve from the optical absorption by using the relation in Fig. \ref{fig:Malpha}.
}
\end{figure*}
The light blue color of liquid oxygen is due to the optical absorption of the O$_2$ molecules, that is closely related to its magnetism.
Thus, optical spectroscopy is a useful technique to detect the LLT.
Figure \ref{fig:opt} summarizes typical results of magneto-optical measurements of liquid oxygen.
Optical-absorption spectra [Fig. \ref{fig:opt}(b)] and the magnetic-field profile [Fig. \ref{fig:opt}(a)] are obtained as a function of time.
The results do not show significant hysteresis.
The magnetic-field dependence of the absorption spectra is shown in Fig. \ref{fig:opt}(c).
The spectra from bottom to top correspond to the results from 0 to 140 T in steps of 10 T.
Each curve is shifted by 0.1 for clarity.

The observed absorption peaks at 1.98 and 2.15 eV are called bimolecular absorption \cite{Uyeda1988,Litvinenko68,Bhandari73,Gaididei75,Gaididei76,Nomura2013,Nomura14}.
Here, two O$_2$ molecules with triplet ground state ($^3 \Sigma ^-_g$) are excited to the singlet state ($^1 \Delta_g$) simultaneously as ${^3 \Sigma ^-_g}{^3 \Sigma ^-_g} \rightarrow {^1 \Delta_g}{^1 \Delta_g}$.
This optical-dipole transition is a forbidden process for one O$_2$ molecule, because the spin-quantum number changes.
This optical transition is allowed only for O$_2$ dimers forming a singlet state.
Strong magnetic fields suppress the singlet-dimer formation and decrease the absorption intensity \cite{Uyeda1988,Nomura2013,Nomura14}.
Our experimental results clearly detect the decrease of the absorption intensity for both peaks (0-0) and (0-1), where the index (0-$\nu$) indicates the $\nu$-th vibronic replica.
The slight change of the absorption peak width of the (0-1) peak is due to the orbital Zeeman effect of the excited state ${^1 \Delta_g}{^1 \Delta_g}$ \cite{Nomura14}.
In the following, we discuss the absorption amplitude that is related to the magnetic ground state of liquid oxygen.

Figure \ref{fig:opt}(d) shows the normalized absorption amplitude of the (0-0) and (0-1) peaks as a function of magnetic field.
The absorption amplitude is obtained from the area of the absorption peak, and normalized to the zero-field value.
By repeating the measurement at different temperatures and maximum fields, the error bar is estimated.
The difference between the (0-0) and (0-1) peaks is within the shown error of the measurement.
The observed hysteresis is considered to be extrinsic, probably related to the movement of liquid oxygen or inhomogeneous pressure increase in the cell.

Uyeda {\it et al.} proposes that the bimolecular absorption intensity is proportional to the statistical probability to form singlet dimers, and the energy levels can be approximated by the spin-dimer Hamiltonian \cite{Uyeda1988}.
The statistical analysis gives the normalized absorption as
\begin{equation}
\frac{\alpha(H)}{\alpha(0)}=
\frac{5{\rm e}^{6I}+3{\rm e}^{2I}+1}
{2({\rm e}^{6I}+{\rm e}^{2I}){\cosh}G+2{\rm e}^{6I}\cosh2G+1+{\rm e}^{2I}+{\rm e}^{6I}}.
\label{eq:absAF}
\end{equation}
Here, $G$ and $I$ are the Zeeman and exchange energy, respectively, normalized by the thermal energy as
\begin{equation}
G=g{\mu}_{\rm B}HS/k{}_{\rm B}T,\ \ I=J/k{}_{\rm B}T.
\label{eq:GI}
\end{equation}
From that, one obtains the spin polarization
\begin{equation}
\langle S{}_{\rm z}\rangle
=\frac
{2{\rm e}^{6I}\sinh2G+({\rm e}^{6I}+{\rm e}^{2I})\sinh{G}}
{2({\rm e}^{6I}+{\rm e}^{2I}){\cosh}G+2{\rm e}^{6I}\cosh2G+1+{\rm e}^{2I}+{\rm e}^{6I}}.
\label{eq:mgnAF}
\end{equation}
Here, the magnetization is obtained as $M=g\mu_\mathrm{B}\langle S{}_{\rm z}\rangle$.
A detailed derivation is presented in the Supplemental Material \cite{supple}.
In Fig. \ref{fig:opt}(d), the calculated absorption intensity is plotted using the parameters $J/k_\mathrm{B}=-15$~K and $T=79$~K.
The experimental results are larger than the calculated value, but the difference at the top of the field is within the experimental error.
The difference might come from the pressure increase due to the magnetovolume effect or magnetocaloric effect \cite{Nomura2017MCE}.

\begin{figure}[ptb]
\centering
\includegraphics[width=7cm]{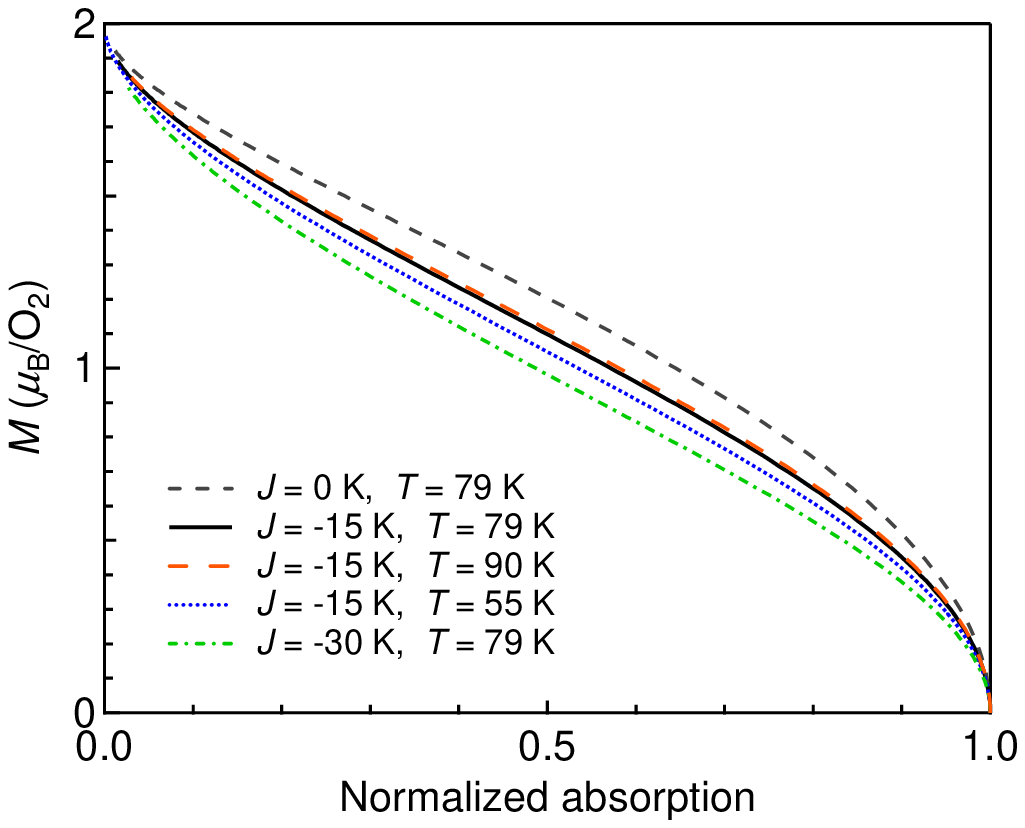}
\caption{\label{fig:Malpha}
Relation between the bimolecular absorption intensity and the magnetization calculated  for various parameters by using Eqs. (\ref{eq:absAF}) and (\ref{eq:mgnAF}).
Exchange and temperature parameters are shown in the legend.
}
\end{figure}

By using Eqs. (\ref{eq:absAF}) and (\ref{eq:mgnAF}), the magnetization can be estimated from the optical-absorption intensity.
Figure~\ref{fig:Malpha} shows the magnetization as a function of normalized absorption for several parameter sets.
We note that this relation does not change significantly in the paramagnetic limit for $|J/k_\mathrm{B}|\ll T$.
This condition is usually satisfied for liquid oxygen (90.2--54.4 K).
Therefore, the conversion from optical absorption to magnetization gives reasonable estimates of the magnetic state.
The estimated magnetization using $J/k_\mathrm{B}=-15$~K and $T=79$~K is shown in Fig. \ref{fig:opt}(e).
Here, we note that the estimated magnetization only reflects the information of local O$_2$-O$_2$ dimers, and does not need to be the same as the bulk magnetization.
Nevertheless, the estimated magnetization is in line with the results obtained by the pickup-coil technique (Fig. \ref{fig:mag}).
The magnetization increases linearly even at 143 T without showing any tendency of saturation.

\section{Conclusion}
We presented experimental results on liquid oxygen under ultrahigh magnetic fields.
The ultrasound results, which reveal a drastic increase of sound attenuation, strongly suggest a nearby structural instability of liquid oxygen induced by external magnetic fields.
We relate this to the precursor of a field-induced LLT.
In the search for the proposed LLT, we measured ultrahigh-field properties of liquid oxygen (sound velocity, volume, magnetization, and optical absorption) up to 180 T.
All these results suggest that the proposed LLT does not exist in this field range, and that the magnetization saturates only at higher magnetic fields.
Experiments at even higher fields using the flux-compression technique \cite{Nakamura18} are necessary to observe the saturation, possibly along with the proposed LLT.

The absence of the LLT up to 180 T is a surprising result, because the magnetic field strength is already beyond the energy scale of the magnetic interaction $\Theta \sim -45$ K.
Issues that might extrinsically inhibit the observation of the LLT are first, the pressure increase of liquid oxygen due to the magnetovolume effect.
Because of the $\mu$s timescale of the STC experiments, the pressure increase inside the sample cell cannot be avoided during the experiment.
When the pressure increases, the exchange interaction is enhanced, and the LLT field should increase as well.
However, we note that the density of liquid oxygen cannot become larger than the zero-field value because the liquid oxygen only expands in the closed cell.
Second, the temperature of the liquid oxygen might increase during the experiment because of the magnetocaloric effect.
For a paramagnet, the magnetic entropy rapidly changes near the saturation field of magnetization.
Therefore, the temperature change could be much larger than the reported value ($\sim +1$~K at 50 T) \cite{Nomura2017MCE}.
At increased temperatures, the anomaly related to the LLT might become too weak.
Third, the resolution of the STC experiments might not be sufficient.
This is related to the pressure increase during the pulse.
An inhomogeneous pressure increase perturbs the measurement at equilibrium conditions and results in the observed hysteresis.
If the anomaly is smaller than the noise level including hysteresis, we might overlook the signal related to the LLT.
Further investigations are needed to answer the question, if a field-induced LLT is possible or not.

\section*{acknowledgments}
We thank O. Yamamuro, T. Oda, and M. Obata for fruitful discussions.
We acknowledge the support of the HLD at HZDR, member of the European Magnetic Field Laboratory (EMFL), the BMBF via DAAD (project-id 57457940), and the DFG through the W\"urzburg-Dresden Cluster of Excellence on Complexity and Topology in Quantum Matter--$ct.qmat$ (EXC 2147, project No. 390858490).
T. N. was supported by a Grant-in-Aid for JSPS Fellows.
This work was partly supported by JSPS KAKENHI, Grant-in-Aid for Scientific Research (Nos. 16H04009, 19K23421, 20K14403) and JSPS Bilateral Joint Research Projects (JPJSBP120193507).

\end{document}